\documentclass[twocolumn,prl]{revtex4}% Physical Review L
\usepackage{graphicx}
\usepackage{mathptmx} % postscript fonts
\usepackage{natbib}

\begin{document}
%\hfill BNL-xxx       % for bnl number

\title{
Scaling and correlations in the dynamics of forest-fire occurrence
}
\author
{
\'Alvaro Corral$^\dag$\email{alvaro.corral at uab.es}, 
Luciano Telesca$^*$, and Rosa Lasaponara$^*$
}
\affiliation{
$^\dag$%
%Grup de F\'\i sica Estad\'\i stica,
Departament de F\'\i sica, Facultat de Ci\`encies, 
Universitat Aut\`onoma de Barcelona,
%Edifici Cc, 
E-08193 Bellaterra, Barcelona, Spain \\
$^*$%
Istituto di Metodologie per l'Analisi Ambientale, CNR, C. da S. Loja,
85050 Tito (PZ), Italy
}
\date{\today}

\begin{abstract}
{
Forest-fire waiting times, defined as the time between 
successive events above a certain
size in a given region, are calculated for Italy. 
The probability densities of the waiting times
%%%defined essentially as the box histogram
%%%normalized by the size of each box and by the number of recurrences, 
are found to verify a scaling law, despite that fact that
the distribution of fire sizes is not a power law.
The meaning of such behavior in terms of the possible self-similarity of
the process in a nonstationary system is discussed. 
We find that the scaling law arises as a consequence
of the stationarity of fire sizes and the existence of a non-trivial 
``instantaneous'' scaling law, sustained by the correlations
of the process.
}
\end{abstract}

%\pacs{
%???
%}% PACS, the Physics and Astronomy Classification Scheme.

\maketitle

In the last years, many natural hazards,
like earthquakes, volcanic eruptions, 
landslides, rainfall, solar flares, etc.,
and other similar-in-spirit phenomena
in condensed-matter physics
have been shown 
to be characterized by a power-law distribution of event 
sizes, over many orders of magnitude in some cases
\cite{Bak_book,Turcotte_soc,Malamud_hazards,Sethna_nature}. 
This kind of distribution has profound implications 
for the nature of these phenomena, as it indicates that
extreme events do not constitute a case
separated from the smaller, ordinary ones;
rather, 
the events are generated by a mechanism that operates
in the same way for all the different scales involved, 
and a characteristic size of the events cannot be defined.
In this way, a reasonable question such as
``which is the typical size of the earthquakes 
in this region?'' is impossible to answer.

Comparison with simple self-organized-critical (SOC) 
cellular-automaton models 
suggests that the events that define the dynamics in these %% kind of 
phenomena consist of a small instability or excitation
that propagates as a very rapid chain reaction or avalanche through a 
medium that is in a very particular state, similar
to the critical points found at continuous phase transitions
in condensed-matter physics 
\cite{Bak_book,Turcotte_soc,Sornette_critical_book,Hergarten_book}.
The dissipation produced by each avalanche would act as 
a feedback mechanism that 
balances a slow energy input and
maintains the system close to the critical state.

Of special interest is the case of forest fires,
for which cellular-automaton models 
yielded a power-law behavior for the distributions of burned areas
(which are a measure of the size of the events),
and showed the previous mechanisms at work
\cite{Bak_forest_fire,Drossel}; 
curiously, it was not 
until much later that Malamud {\it et al.} observed
power law distributions
for real forest fires, with exponents around 1.4
for the probability density (i.e., non-cumulative
distribution) of fire sizes
\cite{Malamud_science,Turcotte_physA,Malamud_fires_pnas}.

Nevertheless, this issue is %%%certainly debatable, 
still open,
as other studies with different data
do not agree with a simple power-law behavior:
Ricotta {\it et al.} \cite{Ricotta_99} postulated that 
fires of large sizes, due to negative economic and social effects, 
are reduced by the massive human intervention; 
therefore, less than expected large fires occur,
leading to an increase (in absolute value) 
of the power-law exponent in that regime. 
Reed and McKelvey \cite{Reed_02}, using the concept of 
extinguishments growth rate, presented a four-parameter 
``competing hazards'' model providing the overall best fit. 
In a subsequent paper, Ricotta {\it et al.} \cite{Ricotta_01} 
have observed that a multiple power-law behavior, 
denoted by the presence of different power-law ranges 
delimited by cut-offs, is due to dynamical changes, linked to 
``more or less abrupt changes in the landscape-specific 
process-pattern interactions that control wildfire propagation, 
rather than statistical inaccuracies''. 
Therefore, the appearance of different size ranges 
with different power-law exponents can be accounted for different 
dynamics, involving topographic, climatic, vegetational, and human factors 
\cite{Telesca_05}.

In addition to the size of the events,
the dynamics of event occurrence is of fundamental interest.
Notably, the temporal properties of 
some popular SOC cellular-automaton models
were shown to be described by a trivial Poisson process,
which 
prevented progress in this aspect %% of these phenomena,
until very recently, when it has been concluded that 
this picture is not appropriate \cite{Paczuski_btw};
in parallel, it has been found 
that real systems show a very rich behavior in time,
%%the real systems are richer than the models,
with power-law distributions for the time between 
events and/or scaling laws for these distributions
\cite{Boffetta,Corral_prl.2004,Baiesi_flares}.
%%For the case of earthquakes, for example, the
%%probability density $D_w(\tau)$ for the time interval between 
%%consecutive earthquakes $\tau$ in a certain area
%%verifies
%%$$
%%D_w(\tau) = R(s_c) F(R(s_c) \tau),
%%$$
%%when different windows $w$ in the size of events are considered, 
%%with $R(s_c)$ the rate of occurrence of events for each
%%window and $F$ a scaling function independent of the window $w$.
%
The existence of such scaling laws %% for the time-interval distribution
has implications no less deep than the fact of having a power-law
distribution of event sizes, although they have been much less studied:
(i) the scaling law reflects the fact that the occurrence of large events
mimics the process of occurrence of smaller ones 
(and this behavior is not implicit in the distribution of event sizes),
thus allowing to model the scarce big events on the basis of the abundant small ones;
%and therefore the more abundant small events can be used
%as a model for the more scarce big ones.
(ii) the scaling law is the signature of the invariance of
the process under a renormalization-group transformation,
which strengths the links between natural hazards and critical 
phenomena \cite{Corral_prl.2005}.

\begin{figure*}
\centering
\includegraphics[width=3.5in]{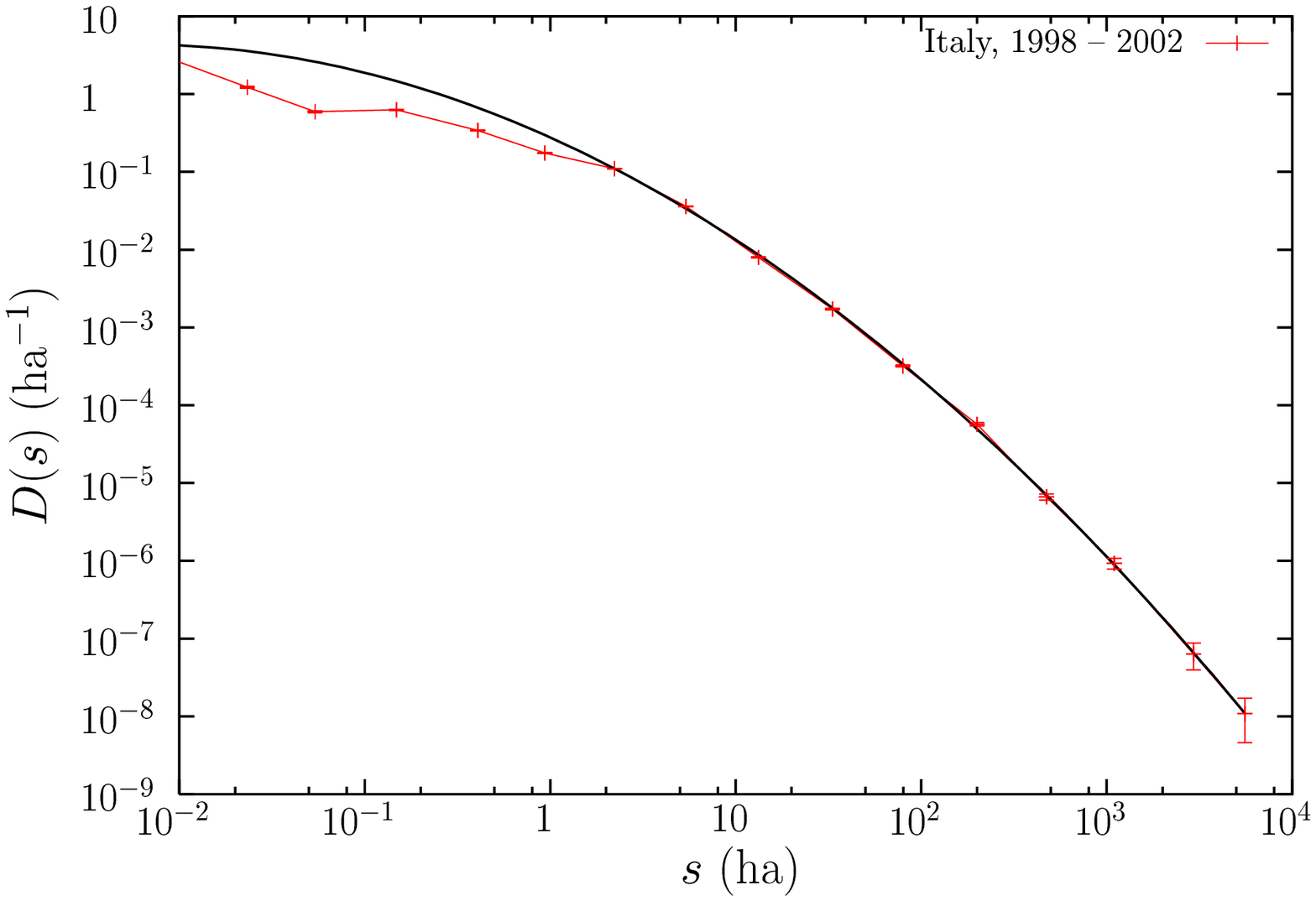}
\caption{
%(color online) 
%%Top curve:
Probability density of fire sizes in Italy, from 1998 to 2002 (included).
%bottom curves (shifted downwards): the same fires separated for the different seasons
%(spring, summer, autumn, and winter), in order to show the
%stationarity of fire sizes.
The error bars are calculated for one standard deviation in the number of counts.
The fit is the lognormal distribution whose parameters are given in the text.
%%several power laws are shown for comparison.
\label{Dm}
}
\end{figure*}

We study in this paper the relation between the
temporal properties
of forest-fire occurrence and the size of the fires,
using the 
AIB (Archivio Incendi Boschivi) fire catalog compiled by 
the Italian CFS (Corpo Forestale dello Stato) for all Italy
\cite{Italy_fire_catalog},
covering the years 1998--2002 (included)
and containing 36821 fires.
In order to characterize the overall behavior,
we measure for the whole catalog
the probability density of the burned areas $s$, 
defined as 
\begin{equation}
D(s) \equiv \frac{\mbox{Prob}[s \le \mbox{ area }
< s + ds]}{ds},
\label{distribution}
\end{equation}
where $ds$ is the bin size (small enough
to sample almost continuously $D(s)$ 
but large enough to guarantee statistical 
significance); 
the resulting shape for $D(s)$ is shown in Fig. \ref{Dm}.
Although a power law could be fit to the data,
it is clearly seen that the curve is continuously
bending downwards, which is the characteristic
of a lognormal distribution,
\begin{equation}
D(s)=
\frac C {\sqrt{2\pi} \, \sigma s} \, \exp\left(-\frac
{\left(\ln s - \mu\right)^2}{2\sigma^2}\right )
\propto  
%{s^{-\left(1+\frac 1 {2\sigma^2} \ln  \frac s {e^\mu}\right)}},
%$$
%$$
%= {s^{-\left(1+\  \ln (s /e^\mu)/(2\sigma^2)\right)}},
%= 
\left( \frac {e^\mu} s \right)^{1+\ \frac{ \ln (s /e^\mu)}{2\sigma^2}},
\label{lognormal}
\end{equation}
with $\mu$ and $\sigma$ the mean and standard deviation of $\ln s$,
and $C$ a correction to normalization due to the fact that
the fit is not valid for all $s$. 
In this way, for each $2\sigma^2$ that $\ln s$ is away from $\mu$
the exponent of the previous pseudo-power law increases in one unit
(in other words, each decade $s$ is above $e^\mu$ increases 
the exponent in $\ln 10/ (2\sigma^2)$).
When $s$ is measured in hectares (ha), 
the results of the best fit yield 
$\mu=-0.35$, $\sigma=9.5$, and $C=6.7$;
this fit holds not only for the full data
but it can be verified that
also describes smaller parts of the country and
shorter periods of time.
%%Figure \ref{Dm} also shows $D(s)$ for the different seasons
%%of the year, leading, remarkably, to the stationarity of fire sizes.
In any case, we have no means to conclude 
if the deviation from a power-law behavior 
is due to human extinction efforts
or to the territorial characteristics
of a high-populated country.

%An alternative way to characterize fire size is 
%using the fire duration. The probability density of
%this quantity is shown in ??? and is well described
%again by a lognormal distribution, 
%this time with parameters ...???
%These values are in agreement with a power-law
%relation between areas and durations...

From the distribution of sizes,
knowing the total number of events, 
it is possible to calculate
the mean waiting time (or recurrence time) 
for events above a certain size $s_c$
\cite{Malamud_fires_pnas};
however, looking at the individual values of the waiting times
one sees that they are broadly distributed and
therefore the mean values
are not very informative about the dynamics of the process;
so, in order to investigate the temporal properties of fire occurrence
it is necessary to look at the whole waiting-time distribution.
To be precise, the procedure is as follows:
once a spatial area, a time period,
and a minimum event size, $s_c$, are selected,
%%a space-time-size window $w$
%%for which the fires there can be considered as 
the fire history is described as a simple point process,
$\{t_0, t_1, t_2 \dots\}$, where $t_i$ denotes 
the time of occurrence of fire $i$.
%% in window $w$.
For this process, the set of waiting times, defined as
the time intervals between consecutive events,
is obtained straightforwardly as 
$
\tau_i \equiv t_i - t_{i-1}.
$
%%%and from here the waiting-time probability density
%%%in window $w$, $D_w(\tau)$, may be obtained.
%%in a way analogous to $D(s)$.
%
%%or noncumulative distribution, is obtained as
%%$$
%%D_w(\tau) \equiv \frac{\mbox{Prob}[\tau \le \mbox{ waiting time }
%%< \tau + d\tau]}{d\tau},
%%$$
%%where $d\tau$ is an interval of waiting times whose
%%length has to be taken appropriately in order 
%%that the bins contain sufficient statistics but
%%without hiding the continuous variation of the distribution.
%%A simple solution is to take $d\tau_n = b^n(b-1)$, 
%%with $n=0,1\dots$ and $b > 1$,
%%which gives equally-sized bins in a logarithmic scale.
%%The probability Prob is estimated just as the ratio of
%%waiting times within an interval to the total number
%%of waiting times ($N_w-1$).
%
%
%
Important insight into the nature of the process may 
be obtained by considering $s_c$ not as a constant
but as a variable parameter \cite{Bak.2002,Corral_prl.2004}, 
and then, the waiting-time probability density for the 
selected window, defined in the same way as in Eq. (\ref{distribution}), 
will be also considered as a function $D(\tau;s_c)$
of the minimum size $s_c$.

\begin{figure*}
\centering
\includegraphics[width=3.5in]{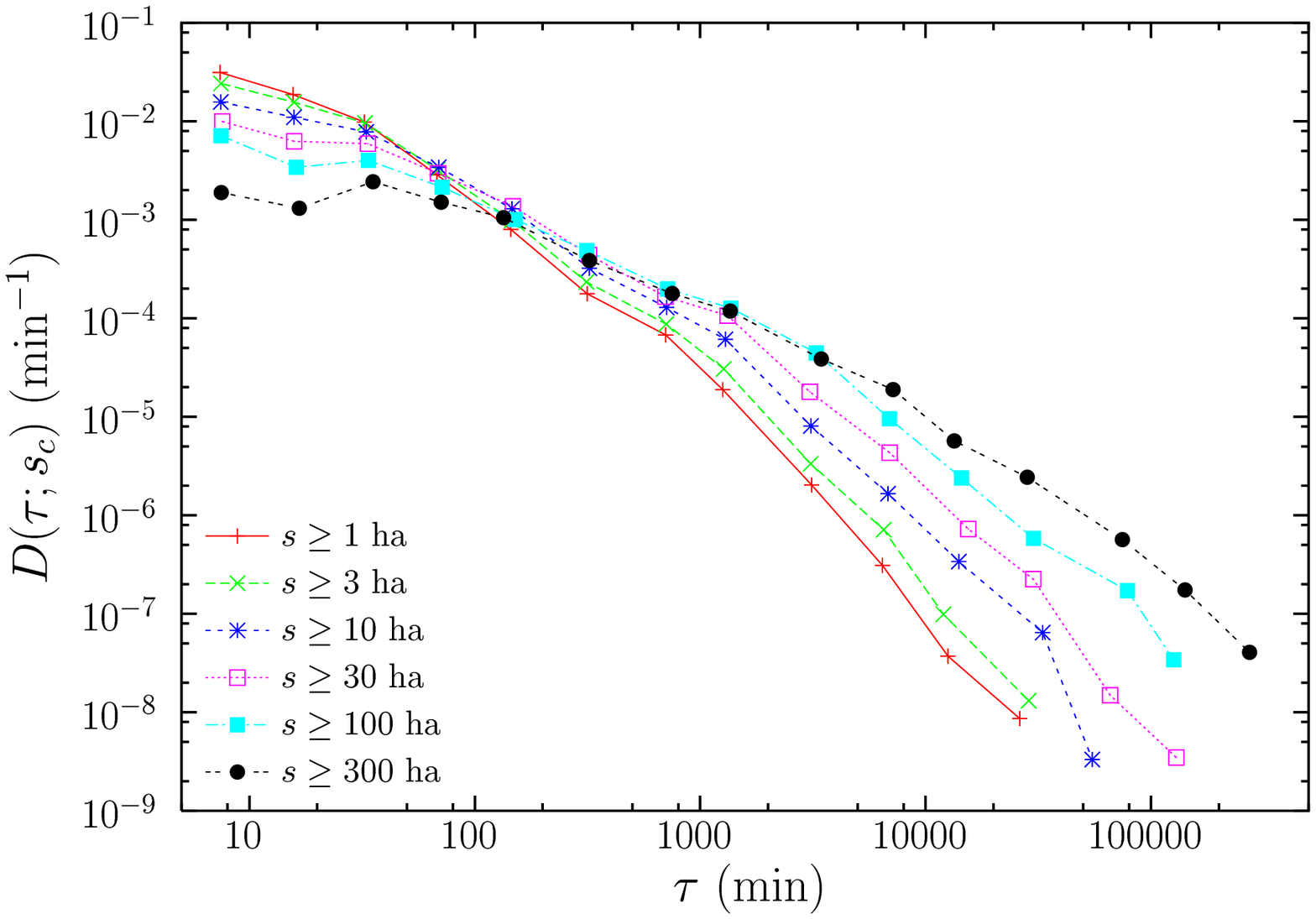}
\includegraphics[width=3.5in]{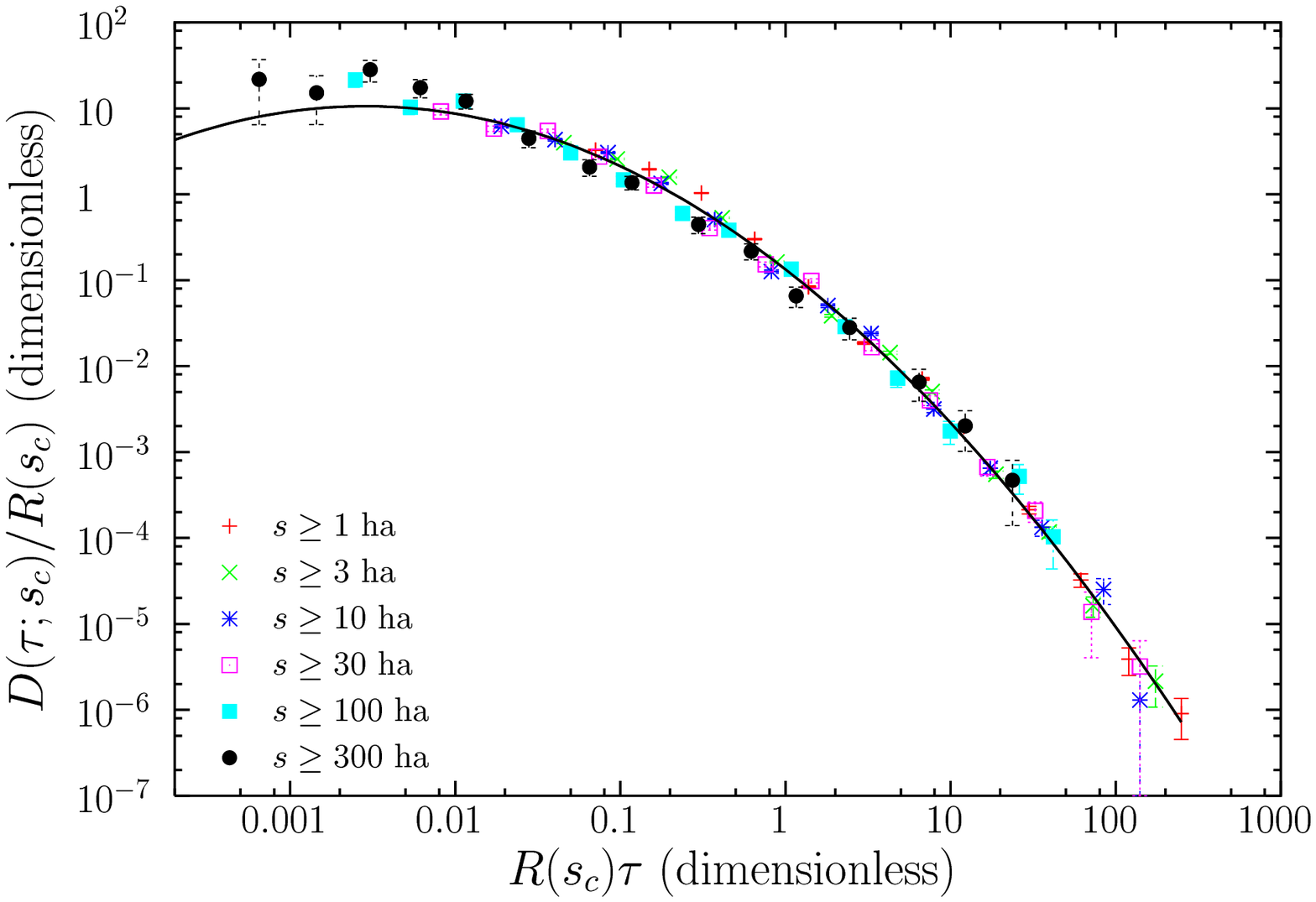}
\caption{
%(color online)
(a) Probability densities of fire waiting times for Italy, from 1998 to 2002
(included) for different minimum burned areas ($s_c=1$ ha to $s_c=300$ ha);
waiting times smaller than 5 min are not plotted.
The slope of the tail seems to decrease with increasing $s_c$.
(b) The previous densities after rescaling by the mean fire rate,
adding error bars (corresponding to one standard deviation). 
The data collapse indicates the existence of a scaling law (see text)
and allows a unified description of the shape of the density, 
in terms of a lognormal distribution, rather than as a power law.
%% ojo!!! El fit que se muestra no es el del texto, 
%% es sigma = 1.98082681142127, mu = -1.93396278844232, C = 1.06880336955804
\label{Dt}
}
\end{figure*}

For the whole country and the total temporal extension of the catalog
we obtain %%%for $D(\tau;s_c)$ 
the different set of curves displayed in Fig. \ref{Dt}(a).
We might fit a (decreasing) power law for each distribution, but
the exponent would decrease with the increase of the minimum size $s_c$.
Instead, it is more convenient to rescale the distributions 
in order that all of them have the same mean and can be properly compared;
this is accomplished by the scale transformation 
$\tau \rightarrow R(s_c) \tau$ and
$D(\tau; s_c) \rightarrow D(\tau; s_c)/R(s_c)$, 
where $R(s_c)$ is the rate of fire occurrence,
defined as the mean number of fires per unit time with $s\ge s_c$ 
(that is, the inverse of the mean of each distribution).
The results of the rescaling, as shown in Fig. \ref{Dt}(b), 
lead to a collapse of the rescaled distributions into a single 
function $F$, signaling the fulfillment of a scaling law, 
\begin{equation}
D(\tau;s_c) = R(s_c) F(R(s_c) \tau),
\label{scaling}
\end{equation}
in the same way as for several natural hazards
\cite{Corral_prl.2004,Baiesi_flares,Bunde}
and other avalanche-like processes
\cite{Yamasaki,Davidsen_fracture}.

The rescaled  %%%rescaling %% rescaled??
plot unveils more clearly the behavior of the distributions:
instead of different power laws, what we have is a unique shape,
but at different scales.
Again, the apparent continuous decrease of the exponent with the rescaled time,
$\theta \equiv R(s_c) \tau$,
suggest a lognormal shape for $F(\theta)$ as that of Eq. (\ref{lognormal}), 
where now we will use tildes to denote the parameters.
The best fit yields $\tilde \mu =  -2.0$ and $\tilde \sigma= 2.0$,
fixing $\tilde C\equiv 1$.
%%sigma2 = 2.02855366423825;mu2 = -2.00965782373855
Notice that now we have the constraint
that the mean of the rescaled distribution, $\bar \theta$, has to be one;
as $\bar \theta = e^{\mu+\sigma^2/2}$, this leads to $\mu=-\sigma^2/2$.
%% o sea, si C=1, mu=-sigma^2/2

It is remarkable that, unlike earthquakes, solar flares, or fractures
\cite{Corral_prl.2004,Baiesi_flares,Davidsen_fracture,Astrom},
forest fires fulfill a scaling law for the waiting time distributions 
without displaying power-law distribution of event sizes.
We could conclude that we have self-similarity in size-time
without having scale invariance in size alone.
This self-similarity 
%%%can be understood writing a scale transformation in this way: 
means that for the linear scale transformation
$\tau \rightarrow a \tau$ and $s_c \rightarrow b s_c$,  
the value of $b$ which guarantees scale invariance is given by 
$R(b s_c) a = R(s_c)$,
which means that $b$ does not only depend on $a$, as in 
the case of a power-law distribution of sizes,
but it also depends on $s_c$. 
This would be equivalent to define an artificial new size variable 
enforcing that it be power law distributed.
However, although this picture describes a kind
of self-similarity, it is not a sufficient condition.
Indeed, the seasonality of fire occurrence prevents 
self-similarity in size-time: five years of fire occurrence
cannot be equivalent to one year of smaller fires, 
as there is a clear annual modulation in fire occurrence;
nevertheless, for a fixed time window still the small
events are a model for the occurrence of the big ones.

Which is then the origin of the scaling law (\ref{scaling})?
It is not difficult to relate it with the stationarity
of fire sizes 
and with the existence of a scaling law
for the ``instantaneous'' waiting-time distributions.
Indeed, $D(\tau; s_c)$ is a statistical mixture
of those instantaneous waiting-time distributions $D_t(\tau; s_c)$,
which,
when the scale of variations 
of the rate is much larger than the corresponding mean waiting time,
take into account that fire occurrence is not stationary
but change with time $t$; 
if it is only the instantaneous rate
$r(t;s_c)$ (defined as the number of fires per unit time in a small 
time interval around $t$) what determines fire occurrence, we can write
$D_t(\tau; s_c) =D(\tau; s_c | r(t;s_c))$ and then,
$$
D(\tau; s_c) = \frac 1 {R(s_c)} \int_{r_{min}}^{r_{max}}
               r D(\tau; s_c | r) \rho(r; s_c) \, dr,
$$
where $\rho(r; s_c)$ is the density of rates, i.e., the fraction of
the time the rate is in a particular small range of values, 
divided by that range \cite{Corral_Christensen}.
Assuming the stationary nature of fire sizes
(notice that this is not incompatible with 
the nonstationarity of time occurrence), 
this means that
$r(t; s_c)=p r(t;s_0)$, where the fraction $p$ 
is the probability of having a fire larger than $s_c$
knowing that it has been larger than $s_0$, 
$p=\mbox{Prob}[s\ge s_c] / \mbox{Prob}[s\ge s_0]$;
this implies that the density of rates fulfills a scaling law, 
$\rho(r; s_c)=p^{-1}\rho(p^{-1} r; s_0) \equiv p^{-1} g(p^{-1} r)$.
Finally, with the hypothesis that $D(\tau; s_c | r)$ verifies as
well a (instantaneous) scaling law, $D(\tau; s_c | r) = r f (r\tau)$, we get
$$
D(\tau; s_c) = \frac 1 {p R_0} \int_{p a}^{p b}
               r f(r \tau) p^{-1} g(p^{-1} r)\, dr,
$$
with $r_{min}= pa$, $r_{max}= pb$, and $R(s_c)=p R(s_0)\equiv p R_0$.
A simple change of variables reveals that $D(\tau; s_c)$
is a function of the form 
$p \tilde F(p\tau) \equiv p\int_a^b x^2 f(p\tau x) g(x) dx$,
which is equivalent to the scaling law (\ref{scaling}).
In other words, if fire occurrence under hypothetical 
stationary conditions verifies a scaling law for 
the waiting times (which in this case would be a reflection of the
self-similarity of the stationary process,
as explained above), 
non-stationary conditions keep that scaling valid
(with a different scaling function) as long as fire size remains
stationary and the rate does not become too small
for this description to be invalid.
[On the other hand, for rates so small that the mean waiting 
time is much larger than the larger scale of variation of 
the rate itself (whose existence is not known), 
the structure of $r(t)$ would become irrelevant
and the waiting-time distribution would tend to the exponential form
characteristic of Poisson processes.]

\begin{figure*}
\centering
\includegraphics[width=3.5in]{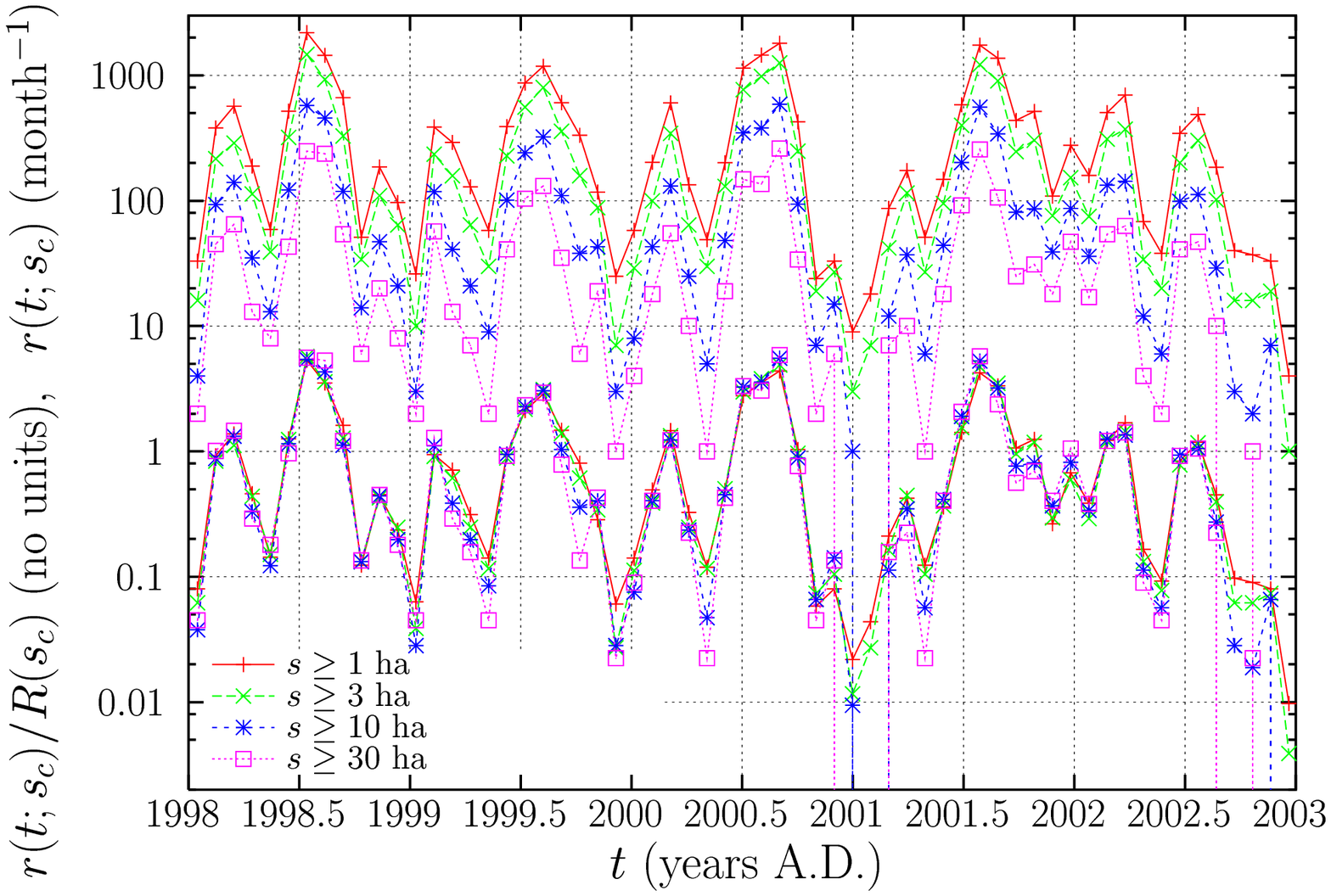}
\caption{
%(color online)
Monthly rates of forest-fire occurrence in Italy 
as a function of time, for events larger 
or equal than $s_c$, with $s_c$ ranging from 1 ha to 30 ha
(top curves).
In addition, the rescaling of the rate by their mean $R(s_c)$
is shown (bottom collapse of curves), 
indicating the stationarity of fire sizes
as well as the nonstationary occurrence in time.
\label{rt}
}
\end{figure*}

\begin{figure*}
\centering
\includegraphics[width=3.5in]{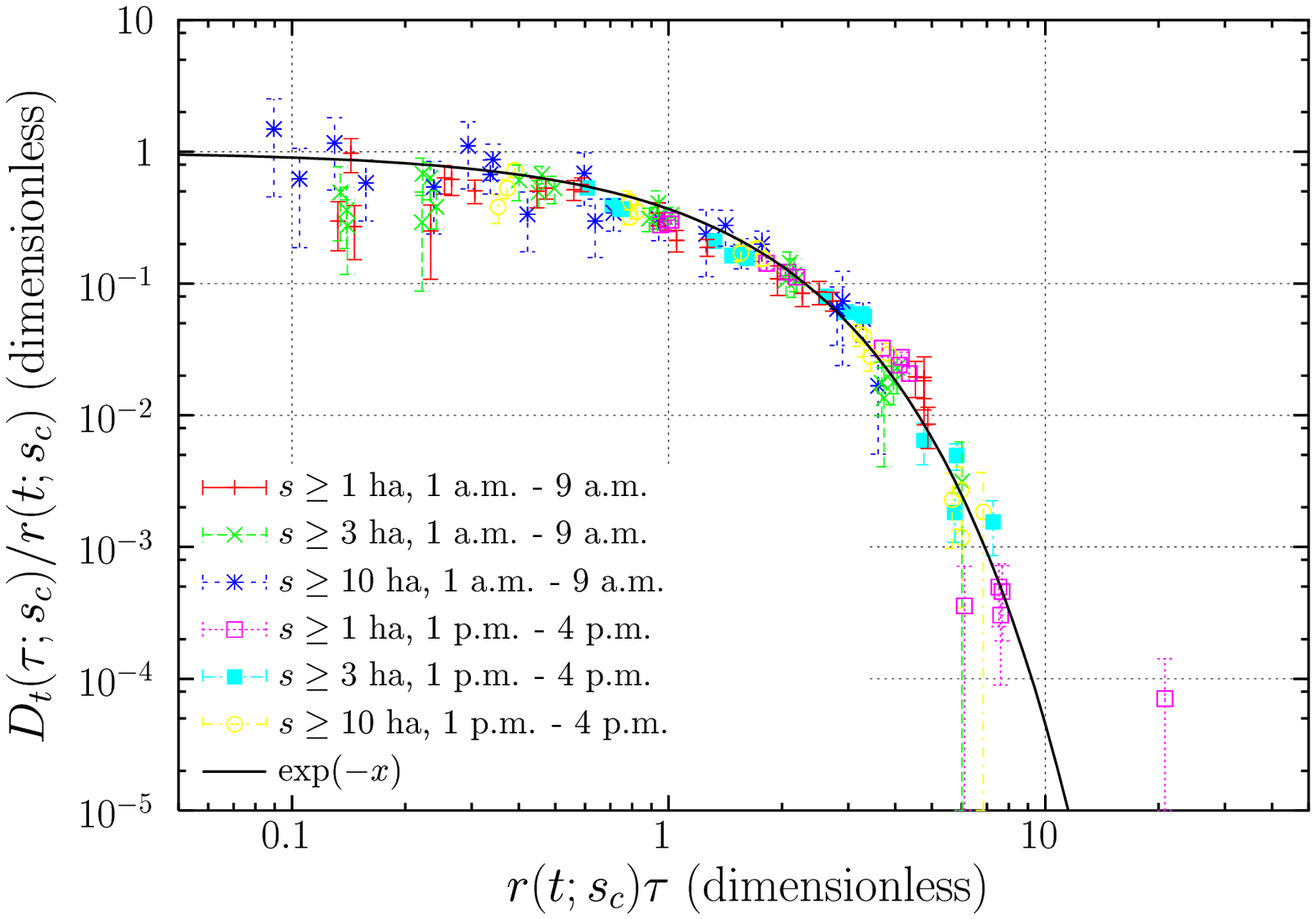}
\caption{
%(color online)
Rescaled distributions of waiting times during annual maximum rate 
periods, corresponding to (in years)
1998.50--1998.65,
1999.60--1999.66,
2000.56--2000.67, and
2001.54--2001.64 (same symbols)
separated for daily minimum rate, 1 a.m. -- 9 a.m.
and maximum, 1 p.m. -- 4 p.m.
Different minimum sizes are used.
The exponential scaling function is shown for comparison.
\label{Dtinstant}
}
\end{figure*}

In order to support our argument for the existence of the scaling law
(\ref{scaling})
we show in Fig. \ref{rt} the stationarity of fire sizes, 
by means of the evolution of $r(t;s_c)$ for different $s_c$,
and how the different curves collapse when they are rescaled by
their mean, $R(s_c)$;
it is also easy to check that the distribution of rates
verifies a scaling law.
The last hypothesis, the scaling of
$D(\tau; s_c | r)$ is more difficult to demonstrate
due to the daily oscillations of $r(t;s_c)$, which makes that
the rate can be considered approximately constant only for a few hours,
corresponding to those of the daily maximum and minimum hazard
(between 1 p.m. and 4 p.m. and between 1 a.m. and 9 a.m.,
respectively).
This short range of variation leads to very low statistics;
nevertheless, for the periods of the year of maximum fire occurrence
(for about one month in the summer) the maximum and minimum 
daily rates are fairly constant for different days, 
which allows to improve the statistics.
The results obtained in this way 
are shown in Fig. \ref{Dtinstant},
although they are not conclusive.
Essentially, they are compatible with 
an instantaneous scaling law, 
with perhaps an exponential instantaneous distribution, 
$D(\tau; s_c | r) \simeq r e^{-r \tau}$, but the statistical
errors are large;
in any case, the hypothesis
of the instantaneous scaling law
cannot be rejected.

If we find an exponential form for 
the instantaneous distributions, does this mean that
the dynamics can be described
by a nonstationary Poisson process?
%%We can test this using a different approach.
%%The simplest model we can propose for fire occurrence in time
%%is a non-stationary Poisson process,
This is the simplest model for nonstationary behavior,
for which the events take place at a rate that does
not depend on the occurrence of the other events, 
as in the simple (stationary) Poisson process, but with the difference
that the rate changes with time (independently on the process, 
we can imagine the rate is related to the meteorological conditions,
not affected by the presence of fire or not).
This leads indeed to exponential instantaneous distributions
(provided the rate is not too small),
although the reciprocal is not true, in general. 
If, in addition,
the size of the events constitutes an independent random process,
this ensures the existence of a scaling law
for the instantaneous distributions (as Poisson processes
are invariant under random thinning plus rescaling, see
\cite{Corral_prl.2005}).
%%We call such a process (nonstationary Poisson plus independent 
%%magnitudes) a nonstationary marked Poisson process.
The nonstationary Poisson process has been recently used for earthquake occurrence,
see Ref. \cite{Shcherbakov}.

A test to verify if a process is of the nonstationary 
Poisson type was introduced by Bi {\it et al.} \cite{Bi}.
One only needs to compute for each $i$ the statistics 
$h_i \equiv 2\tau_{{min }\, i}/(2\tau_{{min }\, i} + \tau_{{neig } \, i})$, 
where $\tau_{{min }\, i}$ is the minimum of $\tau_i$ and $\tau_{i+1}$,
and $\tau_{{neig }\, i}$ is the length of the interval neighbor
of the minimum one opposite to the one used in the comparison, 
i.e., $\tau_{i-1}$ or $\tau_{i+2}$ respectively.
Under the hypothesis we want to test, both 
$\tau_{{neig }}$ and $2 \tau_{{min }}$ 
are independent and
exponentially distributed with approximately the same rate, $r(t)$, 
and therefore it can be shown that
$h$ is uniformly distributed between 0 and 1.

\begin{figure*}
\centering
\includegraphics[width=3.5in]{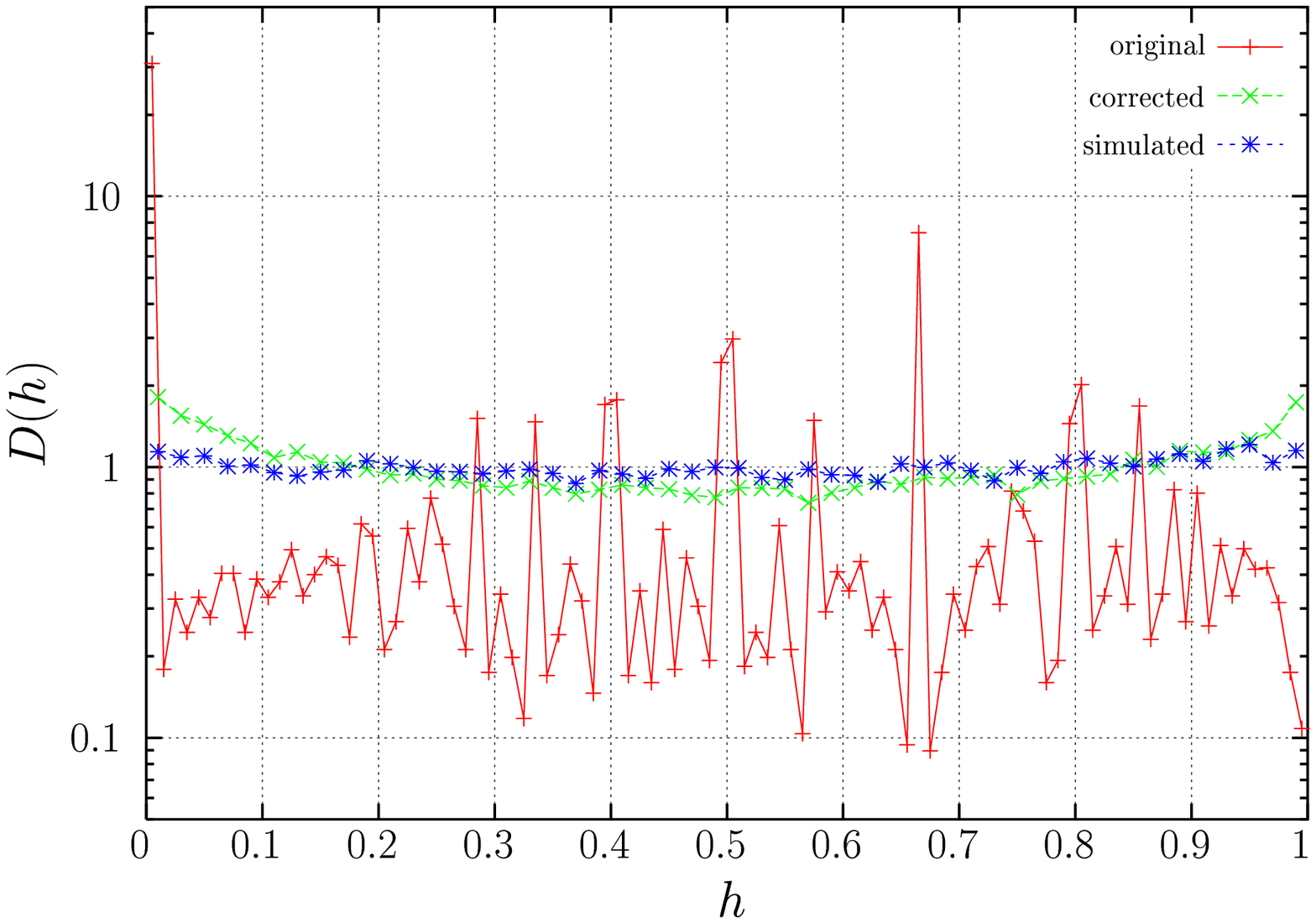}
\caption{
%(color online)
Probability densities of the statistics $h$,
for the original catalog, for the smoothed catalog (corrected),
and for a simulation of a nonstationary Poisson process
with the same rate as the real process
to which the same discretizaion and smoothing procedure has been applied.
\label{bi}
}
\end{figure*}

The application of the test to the fire data yields
catastrophic results, see Fig. \ref{bi}. 
The obtained probability density for
$h$ is far from uniform, with very large peaks for 
precise $h$-values. This is due to the discretization of
fire occurrences in the catalog, which are determined verbally
and therefore rounded mainly in units of 10 or 15 min;
this favors particular values of $\tau$
and there fore of $h$ (2/3, 4/5, 1/2, etc.).
We can correct this effect by the addition of a 
uniform random value between -5 min and 5 min to each 
occurrence time $t_i$, then the peaks in the distribution of $h$ disappear
and its shape gets closer to a uniform one;
however, the difference is significant.
We have verified that the difference is not due to the 
random addition we have performed:
simulation of a non-stationary Poisson process 
where the occurrences are rounded in intervals of 10 min
yields a perfect uniform distribution when this discretization is 
corrected by the uniform random addition just explained (Fig. \ref{bi}).
In consequence, this model does not seem suitable for fire occurrence,
and although the instantaneous distributions are close to exponential
(Fig. \ref{Dtinstant}), 
this is not a sufficient condition to have a nonstationary Poisson
process, as the absence of correlations is equally important
for it.

%This could be an indication of the validity of
%the non-stationary Poisson process to account for the 
%temporal occurrence of fires, in that case, 
%not only the instantaneous distribution has to 
%be an exponential function but also the occurrence of events
%has to be independent on each other.

%%Nevertheless, fire occurrence is a seasonal phenomenon, and the
%%waiting-time distributions we have obtained is an average over
%%different regimes.
%%Figure ??? shows the waiting-time densities for several summer periods, 
%%characterized in addition for a nearly constant rate of occurrence.

\begin{figure*}
\centering
\includegraphics[width=3.5in]{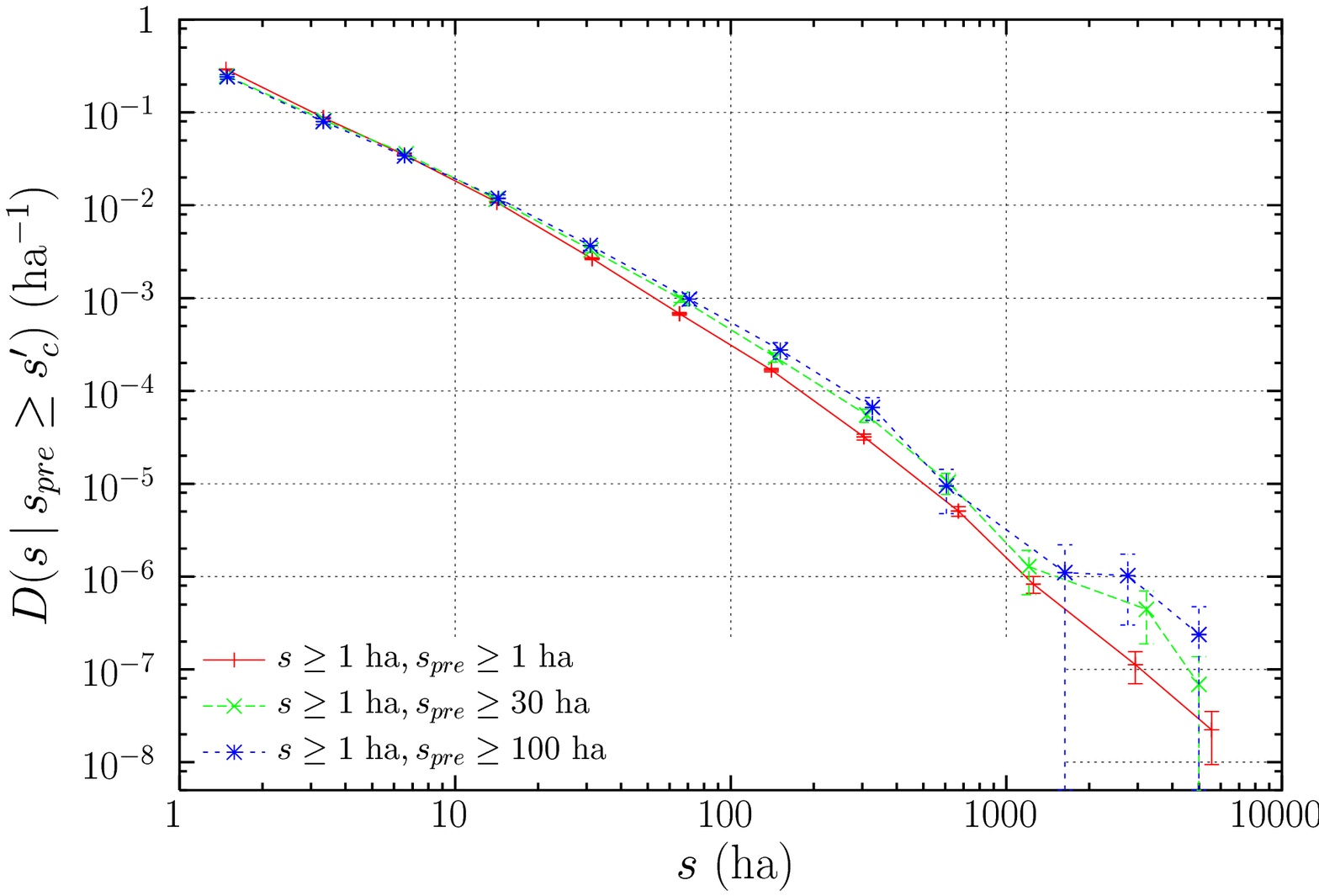}
\caption{
%(color online)
Conditional probability densities of fire sizes in Italy when the previous fire
is larger than $s_c'$. Only fires greater than 1 ha have been taken into account.
The case $s_{pre} \ge 1$ ha corresponds to the unconditioned distribution 
shown in Fig. 1.
The change of the conditional distributions with respect the unconditioned
case is an indication of positive correlations between fire sizes.
\label{Dscond_sold}
}
\end{figure*}

%%Once 
If we reject
the nonstationary Poisson process with %%additional 
independent sizes as a model of fire occurrence, 
the only way to get a scaling law for the instantaneous process
is by means of orchestrated
correlations between sizes and occurrence times
\cite{Corral_prl.2005}.
%
%At a first sight, it may seem that fire occurrence displays strong
%correlations between waiting times, as its rate changes
%dramatically with time, but this correlations are spurious, 
%as they are due to the seasonality of the phenomenon, 
%with much more events in summer than in autumn, for instance.
%A non-stationary Poisson process can account for the main
%features of these temporal variations
%(notice the difference with the behavior of the size of fires,
%which are stationary, as shown in Fig. \ref{Dm}).
%
%%
%
In order to establish the existence of such correlations
we proceed to study conditional size distributions,
defined as in Eq. (\ref{distribution}) but with an additional 
condition for the computation of the probability.
We consider $D(s \,|\, s_{pre} \ge s_c')$, 
which accounts for the size of the events
for which
the size of the immediate previous-in-time event, $s_{pre} $, 
is above a given threshold $s_c'$.
The results in Fig. \ref{Dscond_sold} show that
an increase of $s_c'$ triggers a greater proportion
of large fires, i.e., large fires are followed by large fires.
The dependence of a fire size on the previous size is small 
but significant, unlike to what happens for earthquakes, 
where correlation between their magnitudes has
not been detected \cite{Corral_comment,Corral_tectono}
(nevertheless, for an alternative view see Ref. \cite{Lippiello}).
On the other hand, 
the dependence of waiting times on the size of the
event defining the starting of the waiting period
%%%($pre$ event)
can be measured by $D(\tau ; s_c \,|\, s_{pre} \ge s_c')$ 
%and displayed in Fig. \ref{Dscond_sold}(b).
showing how large fires cause a decrease in the
number of long recurrence times, i.e., those fires
tend to be closer in time to the next fires.
The effect is again small, but clearly detectable,
and in this case has a counterpart for earthquakes, 
where the Omori law for aftershocks implies the same behavior.
%
%%Therefore, we have that a large fire causes the 
%%waiting time to be small, and the size of the next
%%event to be large, 
%

\begin{figure*}
\centering
\includegraphics[width=3.5in]{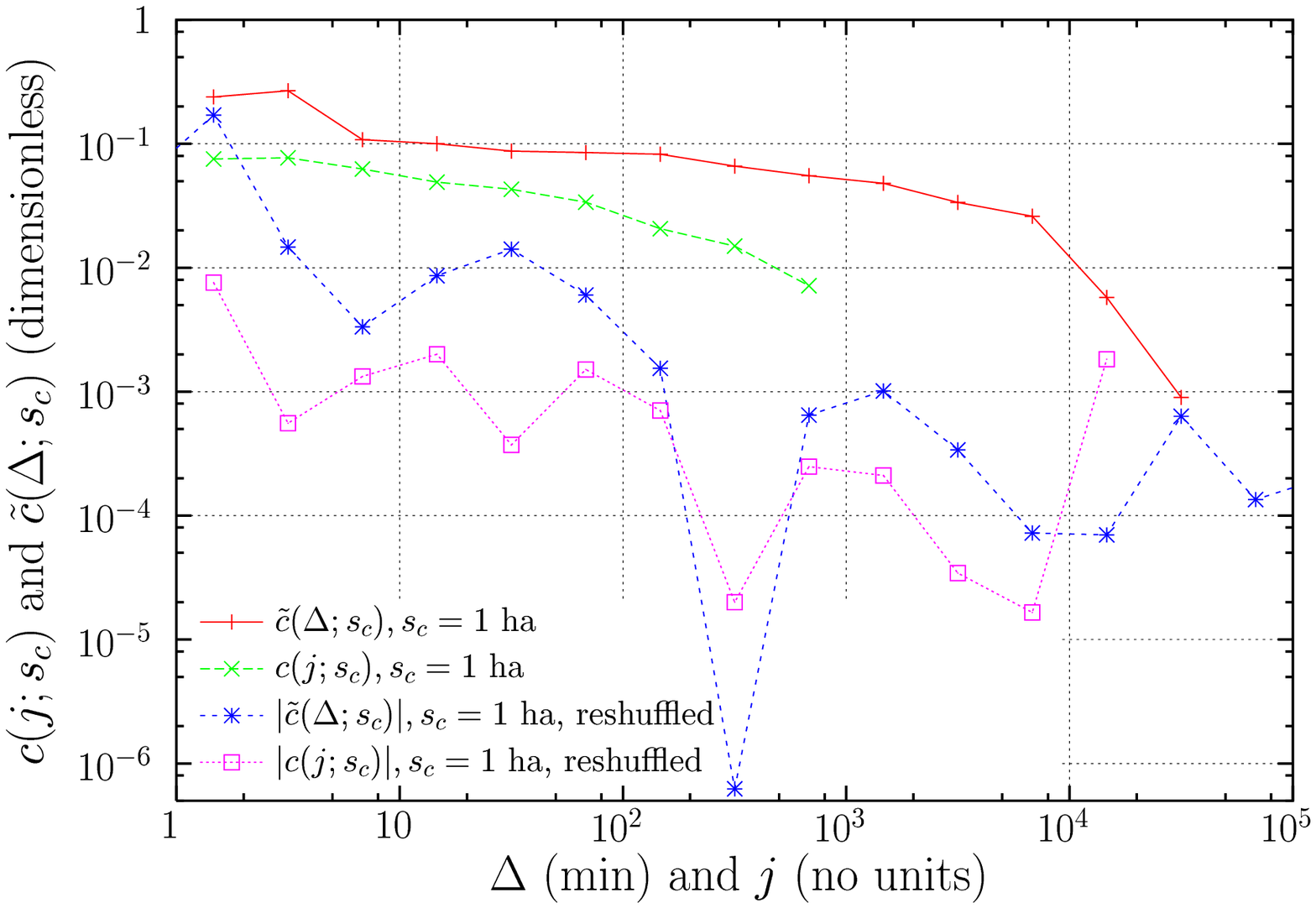}
\caption{
%(color online)
Autocorrelation functions for the Italian catalog
and for a version with reshuffled sizes.
In the latter case it is the modulus of the autocorrelation
what is shown, as the function fluctuates around zero 
and it is equally likely that it is positive or negative.
\label{corr}
}
\end{figure*}

But the correlations between fires are not only with the previous event;
its range can be 
quantified by means of the following auto-correlation function,
$$
c(j; s_c)= 
\left \langle (\log s_i-\bar \ell)(\log s_{i+j}-\bar \ell)
\right \rangle \sigma_\ell^{-2},
$$
where
$\bar \ell$ is the arithmetic mean of the logarithm
of the size (i.e., the logarithm of the geometric mean of the size),
and $\sigma_\ell$ is the standard deviation of the logarithm;
both $\ell$ and $\sigma_\ell$ depend on $s_c$.
Notice that although the process is not stationary,
the stationarity of the size gives sense to the autocorrelation 
function defined in this way.
The results for this function are shown in 
Fig. \ref{corr}, and compared with the same correlation 
function calculated for a reshuffled version of the catalog,
for which the size of the events are randomly permuted,
breaking the correlations between them
(which should yield an autocorrelation function 
fluctuation around zero).
The conclusion is that positive correlations extend significantly
beyond several hundreds of events
(for events of size larger than 1 ha).

More clear is the behavior of the autocorrelation as a function 
of time; as the process is not stationary both functions
are not equivalent. We define
$$
%%c(\Delta; s_c)= \langle (s(t)-\bar s)(s(t+\Delta)-\bar s)\rangle
%%\sigma_s^{-2},
\tilde c(\Delta; s_c)= 
\left \langle (\log s(t)-\bar \ell)(\log s(t+\Delta)-\bar \ell)
\right \rangle \sigma_\ell^{-2},
$$
%%$$
%%cc(\Delta; s_c)= \langle (s(t)-\bar s)(\tau(t+\Delta)-\bar \tau)\rangle
%%\sigma_s^{-1} \sigma_\tau^{-1},
%%$$
where $s(t)$ denotes the size of the fire that
happens at time $t$ 
%%and $\tau(t)$ the waiting time
%%from the event at $t$ to the next event, at $t_{next}$, 
%%i.e., $\tau(t)=t_{next}-t$
(we slightly change notation, for convenience).
The average is taken over all times $t$ and $t+\Delta$ 
for which there are fires, this yields the results of
Fig. \ref{corr}. 
The correlation is again positive, but larger in this case,
suggesting that real time is a better variable to 
describe the evolution of correlations, which extend
for about 10000 min, i.e., roughly 1 week.
It is likely that these correlations are mediated through
the meteorological conditions.

In summary, the dynamics of forest-fire occurrence shows a
complex scale-invariant structure at any time, 
modulated by seasonal and daily variations and orchestrated
by means of broad-range correlations.

%%%N(t)??

%%%D(t) estacionario

%%%D(t) condicionado

%%% Distincion power laws vs scaling laws!!!

%%-- Rescaling para tau > 10 min??

%%-- Y el scaling de los rates!

%%-- Buscar forest fires (wild fires) and climate change

%%The problem is transferred then to determine the origin
%%of the scaling law for the instantaneous distributions.!!!!

%%Leyes de scala para distribuciones condicionales?

%In consequence, we will concentrate on the other kind of correlations:
%i) those between fire sizes only and ii) those between sizes and waiting
%times.!!!!!!

\newpage

\bibliographystyle{unsrt}

\bibliography{../../biblio}

\newpage

\end{document}